\PassOptionsToPackage{dvipsnames}{xcolor}
\documentclass[sigconf,screen]{acmart}

\usepackage{microtype}
\usepackage{array}
\usepackage{enumitem}
\usepackage{footnote}  %put it after xcolor package, otherwise there will be a conflict.
\newcounter{refbug}
\renewcommand{\therefbug}{\#\arabic{refbug}}

\newcommand{\labelbugs}[1]{%
  \phantomsection
  \refstepcounter{refbug}%
  \label{#1}%
  \therefbug}

\newcommand{\refbugs}[1]{%
  \ref{#1}%
}
\AtBeginDocument{%
  \providecommand\BibTeX{{%
    \normalfont B\kern-0.5em{\scshape i\kern-0.25em b}\kern-0.8em\TeX}}}

\renewcommand\footnotetextcopyrightpermission[1]{} % removes footnote with conference information in first column
\setcopyright{none} 
\begin{document}

% \title{Constraints and Bugs of, and Evidence-based Recommendations for Academic Search Engines}
\title{Academic Search Engines: Constraints, Bugs, and Recommendations}

\author{Zheng Li}
\affiliation{%
  \institution{Queen's University Belfast}
  \streetaddress{18 Malone Rd}
  \city{Belfast}
  \state{Northern Ireland}
  \country{UK}
  \postcode{BT9 5AF}
}
\email{imlizheng@gmail.com}
\orcid{0000-0002-9704-7651}

\author{Austen Rainer}
\affiliation{%
  \institution{Queen's University Belfast}
  \streetaddress{18 Malone Rd}
  \city{Belfast}
  \state{Northern Ireland}
  \country{UK}
  \postcode{BT9 5AF}
}
\email{A.Rainer@qub.ac.uk}
\orcid{0000-0001-8868-263X}

\renewcommand{\shortauthors}{Z. Li and A. Rainer}

\begin{abstract}

\textbf{\textit{Background:}} Academic search engines (i.e., digital libraries and indexers) play an increasingly important role in systematic reviews however these engines do not seem to effectively support such reviews, e.g., researchers confront usability issues with the engines when conducting their searches.
\textbf{\textit{Aims:}} To investigate whether the usability issues are bugs (i.e., faults in the search engines) or constraints, and to provide recommendations to search-engine providers and researchers on how to tackle these issues.
\textbf{\textit{Method:}} Using snowball-sampling from tertiary studies, we identify a set of 621 secondary studies in software engineering. By physically re-attempting the searches for all of these 621 studies, we effectively conduct regression testing for 42 search engines. 
\textbf{\textit{Results:}} We identify 13 bugs for eight engines, and also identify other constraints. We provide recommendations for tackling these issues.
\textbf{\textit{Conclusions:}} There is still a considerable gap between the search-needs of researchers and the usability of academic search engines. It is not clear whether search-engine developers are aware of this gap. Also, the evaluation, by academics, of academic search engines has not kept pace with the development, by search-engine providers, of those search engines. Thus, the gap between evaluation and development makes it harder to properly understand the gap between the search-needs of researchers and search-features of the search engines.

\end{abstract}




\maketitle
\pagestyle{plain}

\section{Introduction}

Academic search engines are increasingly important for research in all disciplines.
% have become indispensable and central to higher education and research in all the disciplines. 
Correspondingly, search engine evaluation is significant both to help users (researchers) to select appropriate search engines and to help producers to improve those engines \cite{Vaughan_2004}. (We use the term ``search engine'' to refer to both content providers, such as the ACM Digital Library, and context indexers, such as Google Scholar.)
In the empirical software engineering domain, evidence-based software engineering (EBSE) heavily relies on academic search engines for evidence collection. Evaluations of search engines are sometimes provided through researchers' own experience. For example, it has been reported that different search engines significantly vary in scope, underlying model, user interface, syntax and filtering mechanisms \cite{Dyba_2007,Shakeel_2018}. Such diversity of engine, with the consequential constraints from such diversity, frequently requires that the ``standard'' search string constructed by the researcher has to be modified and adapted for each engine.

Researchers' evaluations are often, but not always, shared as ``side information'' in the secondary studies. Such evaluations are sparse on details, often being more like short experience reports. Although researchers \textit{individually} may report the problems they encounter to the search-engine developers, as suggested by Kruger et al. \cite{Kruger_2020}, there will still be a lack of \textit{holistic} view for the search engine developers to understand the professional users' needs and to improve their products and services. We therefore decided to perform an extensive evaluation of academic search engines, in order to provide systematic feedback and evidence-based recommendations to both researchers and the relevant engine developers.  For the current paper, we focus on the usability of search engines, as usability and its evaluation are widely recognised as important components of search engine development \cite{Vrana_2007,Khoo_2012}.

As with our previous research \cite{Li_2021}, we adopted a snowball sampling technique to identify secondary studies, and obtained 621 valid and unique samples. These 621 studies employed a total of 42 unique digital libraries and indexers. For each of these 621 studies, we re-ran the longest search string in terms of the number of Boolean operators. Each re-run constitutes one ``user interaction''. Considering the diversity in search strings, settings and operations of the 621 studies, each of our user interactions can be viewed as independent test cases for search engine evaluation. Thus, our collection of test cases may be treated as an extensive regression testing, to gain first-hand experience and close observations on search engine usability.\footnote{The detailed testing logs are shared at \url{http://doi.org/10.5281/zenodo.4447488}}
 
This paper reports the results from our regression testing. In addition to listing search engine constraints, we highlight 13 bugs in eight engines. These bugs do not appear to have been reported previously. By comparing the originally published ``user interactions'' with our own experience, we reveal some of the changes in search engines and suggest future directions for the whole search-engine community. Accordingly, this work makes a threefold contribution.
 %Users expect usable and easy to use information systems adapted to their needs and preferences.\cite{Vrana_2007}
 %Accordingly, this paper makes one primary contribution to knowledge and one secondary contribution, and also presents proposals and recommendations to search engine developers and professional users. Our primary contribution is the identification of 13 bugs in 8 search engines. To the best of our knowledge these bugs have not been identified before.
%%%%%%%%%%%%%%%%%%%%% TO BE DISCUSSED!!!!!!!!!!!!!!!!!!!!!!!!!!!!
\begin{itemize}%[leftmargin=*]
\item By fixing the bugs and taking the evidence-based recommendations, the academic search engine developers can improve their products and services to better satisfy the diverse needs of different users. As argued in \cite{Li_2021}, all the stakeholders in the scientific community should take responsibility for improving the research quality rather than purely emphasise the obligation of researchers.
\item Before the search engine developers fix those bugs and relax the constraints, researchers can refer to our testing results to mitigate the user interaction limits. The available workarounds emerged from this work include avoiding EBSE-unfriendly search engines, bypassing the known bugs, and designing search strings that tolerate the existing engine constraints.  
\item By raising the problems of the current search engines, this work can act as a bridge between the testing community and the search engine community. Considering the inefficiency and fatigue effect of using a large number of random test cases, we argue the need of a systematic test case suite for evaluating the engine usability, and our identified bugs can inspire and motive the design of efficient test cases.
\end{itemize}

%The remainder of this paper is organised as follows. Section \ref{sec:RelatedWork} summarises the related work and justifies the innovation of this research. Section \ref{sec:regression} reports our regression testing of the search engine constraints. The identified bugs are separately reported in Section \ref{sec:bugs}. Section \ref{sec:suggestions} elaborates our recommendations for evolving search engines in the future. Section \ref{sec:conclusions} draws conclusions and explains the needs and directions of the future work.

\section{Related Work}
\label{sec:RelatedWork}
With the growth of search engines, researchers have developed various models and frameworks to facilitate evaluating their usability. For example, by summarising the existing discussions about the term ``usability'' in the search engine context, Jeng \cite{Jeng_2005} proposed a usability evaluation model that comprises four dimensions, such as effectiveness, efficiency, satisfaction, and learnability. By focusing on the user interfaces, Lai et al.~\cite{Lai_2014} developed a reference model and identified top five criteria (namely ease of use, searching, language, presentation, and design) for evaluating search engines. By defining the quality components of usability, Nielsen \cite{Nielsen_2012} pointed out another five criteria for usability testing, including learnability, efficiency, memorability, errors, and satisfaction.
Our work only focuses on one evaluation dimension, i.e.~effectiveness, as explained and justified later.

It has been identified that those reference models and criterion frameworks mainly contribute to the theoretical landscape from the researcher's perspective \cite{Li_Liu_2019}, while they offer little handy guidance to practitioners about evaluating the usability of search engines in action. In fact, it is argued that the usability assessment should be an evidence-based practice, via ``carefully observing users interacting with the system in a realistic way'' \cite{Kelly_2014}. Note that the widely employed usability testing methods like questionnaire and survey (e.g., \cite{Xie_2008}) do not satisfy this argument in terms of close observations. In contrast, our research is an empirical study that focuses on the real-time interaction between the user and search engines (i.e.~evidence-based practice), rather than on the user's post-usage feedback.

Some other researchers also emphasised user interactions for academic search engine evaluation, and they further clarified information resource, interface, and tasks as three interaction components \cite{Li_Liu_2019}. However, the aim of their research was to explore the relationships between the user interaction and the search engine evaluation. Differently, our research mainly aims to explore evidence-based recommendations for engine developers to improve their products and services. In particular, although our empirical study replicates search activities of 621 secondary studies, the replication only enables real-time observation on limited users (i.e.~the authors). Therefore, this research reflects only one attribute of usability -- effectiveness (i.e.~how usable) that is concerned with functional completion of user tasks \cite{Buchanan_2009}. There is no doubt that the other non-functional attributes (i.e.~how easy to use) would require observations on the real-time interaction of a crowd of users.

In the context of EBSE, there are early research efforts dedicated to investigating user interactions against academic search engines \cite{Bailey_Zhang_2007,Chen_Babar_2010}. However, these investigations only chose limited engines, and mainly aimed to raise the researchers' understanding and awareness of the inconsistent or unpredictable engine behaviours. A typical example is the development of a caution guide for systematic reviewers, based on the features of five search engines only \cite{Singh_Singh_2017}. To the best of our knowledge, this research is the first investigation that conducts extensive user interaction testing, both in terms of number of tests and the number of search engines tested.
% and urges search engine developers to enhance the literature-intensive research environment.  

 %Gordon and Pathak \cite{Gordon_1999} drafted generic guidelines composed of seven suggestions for search engine evaluation.    

\section{Regression Testing of the Search Engine Constraints}
\label{sec:regression}

\begin{table*}\footnotesize
  \caption{Current Constraints of Academic Search Engines}
  \label{tab:constraints}
  \begin{tabular}{|>{\raggedright\arraybackslash}p{4.8cm} | >{\raggedright\arraybackslash}p{5.8cm} | >{\raggedright\arraybackslash}p{5.8cm} |}
    \toprule
    String Length Constraints & Boolean Operation Constraints & Search Field Constraints\\
    \midrule
    \begin{minipage}[t]{\linewidth}
    \begin{itemize}[leftmargin=*,noitemsep,topsep=0pt]
    \item DBLP has a maximum limit of 127 characters.
    \item Google Scholar has a maximum limit of 256 characters.
    \item IGI Global has a maximum limit of 100 characters.
    \item JSTOR has a maximum limit of 200 characters.
    \item ScienceDirect has a maximum limit of 500 characters.
    \item SpringerLink does not work well with lengthy search strings (about 1800 characters).
\end{itemize}
\end{minipage} 
&
     \begin{minipage}[t]{\linewidth}
     \begin{itemize}[leftmargin=*,noitemsep,topsep=0pt]
    	    \item ACM DL does not support the proximity operators.
            \item Google Scholar does not recognise wildcards and the Boolean operator \texttt{AND} in its Title search.
            \item IEEE Xplore has a maximum limit of seven wildcards in one string.
            \item ScienceDirect does not support wildcard search.
            \item ScienceDirect has a maximum limit of eight Boolean operators per search field.
            \item Web of Science (WoS) supports left-hand wildcard only for the Topic, Title, and Identifying Code searches.
\end{itemize}
\end{minipage} 
& 

\begin{minipage}[t]{\linewidth}
     \begin{itemize}[leftmargin=*,noitemsep,topsep=0pt]
            \item Google Scholar and SpringerLink do not support zonal settings, except for the Title search.
            \item ScienceDirect does not support exclusive abstract search or keywords search.  
            \item ScienceDirect, SpringerLink and Wiley Online Library (Wiley OL) do not support command line search.
            \item SpringerLink's title search is not recommended, as it does not work as expected. 
            \item Wiley OL does not support disjunction between multiple search fields.
\end{itemize}
\end{minipage} \\
    \bottomrule
  \end{tabular}
\end{table*}

% To represent a large scale of real-world usage of academic search engines, 
We sampled 621 EBSE secondary studies and tried replicating the reported user interactions, i.e.~searches. Our first-hand replication trials can be viewed as regression testing of the relevant search engines. The sampling methodology and rationale remain the same as our previous work \cite{Li_2021}. To avoid duplication, and due to publication constraints, we do not repeat them here.  Although we do not report the methodology we do explain, in Section~\ref{sec:bugs}, how to reproduce the bugs we have identified.

After excluding the inapplicable cases (e.g., the retired or inaccessible search engines), we conducted tests on 42 digital libraries and indexing platforms. %\footnote{The 42 digital libraries and indexing platforms are: 1) ACM DL; 2) AIS eLibrary; 3) Australian Education Index; 4) BASE (Bielefeld Academic Search Engine); 5) Blackwell-Synergy; 6) Cambridge University Press; 7) CiteSeerX; 8) CSB (The Collection of Computer Science Bibliographies); 9) DBLP; 10) EBSCO; 11) Embase; 12) Emerald Insight; 13) Emeroteca Virtuale; 14) Engineering Village; 15) ERIC (Education Resources Information Centre); 16) Expanded Academic; 17) Google Scholar; 18) IEEE Xplore; 19) IET Digital Library; 20) IGI Global; 21) InderScience Online; 22) INFORMS PubsOnLine; 23) IOPscience; 24) JSTOR; 25) Kluwer Online; 26) Metapress; 27) Microsoft Academic Search; 28) MIS Quarterly; 29) MIT Press; 30) Oxford Journals; 31) ProQuest; 32) SAGE Journals; 33) Science (AAAS); 34) ScienceDirect; 35) Scopus; 36) SIAM (Society for Industrial and Applied Mathematics); 37) SpringerLink; 38) Taylor \& Francis Online; 39) University of Hertfordshire's Library Search; 40) Wiley OL; 41) WoS; 42) World Scientific.} 
%For the convenience of reporting in this paper, we intentionally ignore the conceptual difference and name those indexing platforms (e.g., Google Scholar) also as DLs. Eventually, we identified a set of user interaction limits due to the DL constraints, and surprisingly, our tests found 13 bugs of eight DLs, as reported in the following subsections.
 We notice that not all the published user interactions are replicable, even by trying to adapt the searches and configurations to the individual interfaces of different search engines. By collecting and analysing the reasons that hindered us from replicating the previous practices, we identify concerning differences, even conflicts, between the researchers' needs and the search engine constraints. 

For example, to maximise the coverage of the relevant primary studies, the de facto guidelines \cite{Kitchenham_Charters_2007,Petersen_2015} for secondary studies generally require exhaustively selecting and combining search terms. This will inevitably result in long search strings. Unfortunately, the well-designed search strings could not be acceptable or executable in practice, due to search engines' string length limits and/or Boolean operation limits. Similar to the Boolean operation limits, the pre-designed search fields on the engine interfaces may also constrain the combinations of search terms. Therefore, we summarise these engine constraints both researchers and practitioners should be aware of, as listed in Table \ref{tab:constraints}.

In particular, we also identified three digital libraries that are generally unhelpful, even disruptive to, systematic reviews:
\begin{itemize}%[leftmargin=*]
    \item AIS eLibrary is not recommended for complex search strings, due to its bug of parentheses interpretation.
        \item CiteSeerX is not recommended in general, due to its unexplainable search behaviours.
        \item DBLP is not recommended in general, due to its unique query syntax and the lack of advanced search.
\end{itemize}
Note that although CiteSeerX is statistically one of the popular digital libraries employed by the existing EBSE researchers (in 10.6\% of, or 66 out of 621, sample studies), we do not recommend it to be used, because of its strange behaviours during the user interaction (see~Bug \refbugs{CiteSeerX}).

%\begin{figure}[!t]
%\centering
%\includegraphics[trim=0 5 0 5,clip]{./Pics/enginesInStudy.pdf}
%\caption{Frequency of search source amounts in one EBSE study.}
%\label{fig_enginesInStudy}
%\end{figure}

\section{Bugs of the Tested Search Engines}
\label{sec:bugs}
Our tests also found 13 bugs in eight engines.
Since ``accurately recording the steps leading to a bug is an absolute must'' \cite{Hamilton_2017}, we use concise demonstrations (presented in boxes) to describe each bug and to show how it can be reproduced. Whilst researchers wait on search engine developers to fix these bugs, researchers need to be aware of, and avoid, these bugs when employing the relevant search engines.

\subsection{The Bugs of ACM Digital Library}
%\label{subsubsec:bugACM}
%\vspace{0.5\baselineskip} 
%\noindent
%\textit{1) The Bugs of ACM Digital Library}.
Although ACM DL offers flexible field-specific search including an Edit Query window, it is not convenient to manipulate complex multi-field search, not to mention that the Edit Query window is invisible by default. Unlike IEEE Xplore, ACM DL's advanced search interface does not enable Boolean operator selection between different fields, i.e.~the logic combination of search fields is fixed by \texttt{AND}. To relax this constraint and adjust search strings in the Edit Query window, users will have to conduct a useless search first to make the Edit Query window visible. Thus, this inconvenience can be viewed as a bug especially from the perspective of new users. By taking the search for \texttt{Title:(software) OR Abstract:(engineering)} as an example, we demonstrate how to reproduce this bug in Bug \refbugs{ACM1}.

\begin{savenotes}
\begin{table}[!ht]\footnotesize
\renewcommand{\arraystretch}{1.5}
%\caption{An Example of a Table}
%\label{table_example}
\centering
\begin{tabular}{|p{0.95\linewidth}|}
\hline
\textbf{Bug \labelbugs{ACM1} Steps to reproduce the bug of multi-field search in ACM DL:} (Last test on July 20, 2022)

\begin{enumerate}[leftmargin=4.5ex,after=\vspace{-1.3\baselineskip}]
\item Go to the interface of ACM DL's Advanced Search (\url{https://dl.acm.org/search/advanced}). 
\item In the Search Within panel, select ``Title'' as the search field and input the search term \texttt{software}.
\item Click the plus icon (i.e.~add search field) next to the search term.
\item Select ``Abstract'' as the added search field and input the search term \texttt{engineering}.
\end{enumerate} \\
Note that there is no chance to use the Boolean operator \texttt{OR}, and the Edit Query window does not exist.\\
\begin{enumerate}[leftmargin=4.5ex,after=\vspace{-1.3\baselineskip},before=\vspace{-1.1\baselineskip}]
 \setcounter{enumi}{4}
\item Click the Search button, and it returns results for \textit{[Publication Title: software] AND [Abstract: engineering]}.
\item Click the Edit Search button and return to the interface of Advanced Search.
\end{enumerate} \\
Note that a View Query Syntax panel appears at the bottom of the interface.\\
\begin{enumerate}[leftmargin=4.5ex,after=\vspace{-1\baselineskip},before=\vspace{-1.1\baselineskip}]
 \setcounter{enumi}{6}
\item Expand the View Query Syntax panel, and the Edit Query window appears and contains \texttt{Title:(software) AND Abstract:(engineering)}.
\item Replace \texttt{AND} with \texttt{OR}, and click the Search button again.
\item The search results are finally correct for \textit{[Publication Title: software] OR [Abstract: engineering]}.
\end{enumerate} \\
\hline
\end{tabular}
\end{table}
\end{savenotes}

When it comes to the single-field search, ACM DL seems to have a hidden bug when dealing with the unnecessary outermost parentheses of search strings. Given a search string enclosed within a pair of outermost parentheses, although the initial search results have nothing wrong, re-searching after editing the string in the Term field (instead of in the Query window) will combine both string versions. The simplified steps to reproduce this bug are specified in Bug \refbugs{ACM2}. It should be noted that the outermost parentheses are indeed redundant for search strings, whereas they are not a syntax error.

\begin{table}[!ht]\footnotesize
\renewcommand{\arraystretch}{1.5}
%\caption{An Example of a Table}
%\label{table_example}
\centering
\begin{tabular}{|p{0.95\linewidth}|}
\hline
\textbf{Bug \labelbugs{ACM2}. Steps to reproduce the bug of outermost parentheses interpretation in ACM DL:} (Last test on July 20, 2022)

\begin{enumerate}[leftmargin=4.5ex,after=\vspace{-1.3\baselineskip}]
\item Open ACM DL's homepage (\url{https://dl.acm.org/}).
\item Input \texttt{(software)} in the default Quick Search field.
\item Click the Search button, and it returns results for: \textit{All: software}.
\item Click the Edit Search button and go to the advanced search interface.
\end{enumerate} \\
Note that the Search Term field shows \texttt{software} without parentheses. \\
\begin{enumerate}[leftmargin=4.5ex,after=\vspace{-1.3\baselineskip},before=\vspace{-1.1\baselineskip}]
 \setcounter{enumi}{4}
\item Replace \texttt{software} with \texttt{engineering} in the Search Term field, remain the Search Term field active and press the Enter key.
\item The results are returned for: \textit{[All: software] AND [All: engineering]}.
\end{enumerate} \\
Note that \textit{[All: engineering]} is expected.\\
\hline
\end{tabular}
\end{table}

In addition, there are inconsistent behaviours of ACM DL's default Boolean operation. By default, ACM DL assumes an \texttt{OR} relationship between search terms. However, there will be clear difference between the search results from an operator-specific string and from the string's alternative version that relies on the default operation. The difference can be demonstrated through the steps in Bug \refbugs{ACM3}.

\begin{table}[!ht]\footnotesize
\renewcommand{\arraystretch}{1.5}
%\caption{An Example of a Table}
%\label{table_example}
\centering
\begin{tabular}{|p{0.95\linewidth}|}
\hline
\textbf{Bug \labelbugs{ACM3}. Steps to reproduce the bug of default Boolean operation in ACM DL:} (Last test on July 21, 2022)

\begin{enumerate}[leftmargin=4.5ex,after=\vspace{-\baselineskip}]
\item Open ACM DL's homepage (\url{https://dl.acm.org/}).
\item Input \texttt{software engineering} in the default Quick Search field.
\item Click the Search button, and it returns ``446,113 Results''.
\item Reopen ACM DL's homepage.
\item Input \texttt{software OR engineering} in the default Quick Search field.
\item Click the Search button, and it returns ``448,696 Results''.
\end{enumerate} \\
\hline
\end{tabular}
\end{table}

\subsection{The Bug of AIS eLibrary}
%\vspace{0.5\baselineskip} 
%\noindent
%\textit{2) The Bug of AIS eLibrary}.
%As a less frequent search source in EBSE, we do not have many chances to test AIS eLibrary in this research. However, we still observed a clear bug in its search engine when dealing with parentheses. 
Although AIS eLibrary offers the +/- buttons to include additional criteria, parentheses can be unavoidable for grouping search terms to satisfy sophisticated logics. Unfortunately, it seems that AIS eLibrary cannot interpret parentheses properly, while there is little information about whether or not parentheses are accepted in its search mechanism. The steps to reproduce this bug are described in Bug \refbugs{AIS}.

%\begin{savenotes}
\begin{table}[!ht]\footnotesize
\renewcommand{\arraystretch}{1.5}
%\caption{An Example of a Table}
%\label{table_example}
\centering
\begin{tabular}{|p{0.95\linewidth}|}
\hline
\textbf{Bug \labelbugs{AIS}. Steps to reproduce the bug of parentheses interpretation in AIS eLibrary:} (Last test on July 21, 2022)

\begin{enumerate}[leftmargin=4.5ex,after=\vspace{-1.3\baselineskip}]
\item Go to the interface of AIS eLibrary's Advanced Search (\url{https://aisel.aisnet.org/do/search/advanced/}).
\item Input \texttt{(software) AND (engineering)} in the field next to All Fields.
\end{enumerate} \\
Note that the search string is immediately interpreted as \texttt{( software AND (engineering )} below the text ``Advanced Search''.\\
\begin{enumerate}[leftmargin=4.5ex,after=\vspace{-1.3\baselineskip},before=\vspace{-1.1\baselineskip}]
 \setcounter{enumi}{2}
\item Click the Search button, and it returns ``0 results''.
\end{enumerate} \\
\hline
\end{tabular}
\end{table}
%\end{savenotes}%%%%%%%%%%%%%%%%%%%%%%%%%%%%%%%%%%%%

\subsection{The Bug of CiteSeerX}
%\label{subsubsec:CiteSeerX}
%\vspace{0.5\baselineskip} 
%\noindent
%\textit{3) The Bug of CiteSeerX}.
According to our trials, CiteSeerX shows various unexplainable search behaviours, which makes it unreliable to support usually complex automated search in EBSE. Furthermore, it is even difficult for us to isolate specific bugs from those strange behaviours. Therefore, we try to merge our observations in a concise way into Bug \refbugs{CiteSeerX}. 

%\begin{savenotes}%%%%%%%%%%%%%%%%%%%%%%%%%%%%%%%%%%%
\begin{table}[!ht]\footnotesize
\renewcommand{\arraystretch}{1.5}
%\caption{An Example of a Table}
%\label{table_example}
\centering
\begin{tabular}{|p{0.95\linewidth}|}
\hline
\textbf{Bug \labelbugs{CiteSeerX}. Steps to reproduce the unexplainable search behaviours of CiteSeerX:} (Last test on July 23, 2022)

\begin{enumerate}[leftmargin=4.5ex,after=\vspace{-\baselineskip}]
\item Go to the interface of CiteSeerX's Advanced Search (\url{http://citeseer.ist.psu.edu/advanced_search}).
\item If Title search for \texttt{software engineering}, it will return ``Your search ... did not match any documents''.
\item If Title search for \texttt{\textquotedbl software engineering\textquotedbl}, it will return ``Results 1 - 10 of 25''.
\item If Title search for \texttt{software OR engin*}, it will return ``Results 1 - 10 of 47''.
\item If Title search for \texttt{software OR engineering}, it will return ``Results 1 - 10 of 1,073''.
\item If Title search for \texttt{(software OR engineering)}, it will return ``Results 1 - 10 of 72,011''.
\item If search for \texttt{(software OR engineering) AND empirical}, it will always return ``Results 1 - 10 of 109'' no matter using which field (even the Author or Venue field).
\end{enumerate} \\
\hline
\end{tabular}
\end{table}
%\end{savenotes}

In fact, its early version (namely CiteSeer) was also identified to have some inconsistent and unexplainable search behaviours \cite{Brereton_2007}. We are afraid that the development of CiteSeerX has made the situation even worse. Thus, we do not recommend CiteSeerX as a search engine for EBSE studies until the aforementioned bug is fixed.   

\subsection{The Bugs of IEEE Xplore}
%\label{subsubsec:bugIEEE}
%\vspace{0.5\baselineskip} 
%\noindent
%\textit{4) The Bugs of IEEE Xplore}.
We experienced two bugs in IEEE Xplore during this research. The first one is when conducting field-specific search, if the search string includes parentheses, the opening parenthesis will be treated as part of its adjacent search term. This can be a hidden problem in the case of long and complex search strings, because users will merely receive ``No results found'' and may not notice the parenthesis issue. A simple demo of this bug is specified in Bug \refbugs{IEEE1}.

\begin{table}[!ht]\footnotesize
\renewcommand{\arraystretch}{1.5}
%\caption{An Example of a Table}
%\label{table_example}
\centering
\begin{tabular}{|p{0.95\linewidth}|}
\hline
\textbf{Bug \labelbugs{IEEE1}. Steps to reproduce the bug of opening parenthesis interpretation in IEEE Xplore:} (Last test on July 27, 2022)

\begin{enumerate}[leftmargin=4.5ex,after=\vspace{-1.3\baselineskip}]
\item Go to the interface of IEEE Xplore's Advanced Search (\url{https://ieeexplore.ieee.org/search/advanced}).
\item Input \texttt{(software AND engineering)} in the Search Term field.
\item Select ``Abstract'' as the search field. In fact, any search field can be used for this demonstration including the default ``All Metadata''. 
\item Click the Search button, and it returns ``No results found for (\textquotedbl Abstract\textquotedbl:(software AND \textquotedbl Abstract\textquotedbl:engineering))''.
\end{enumerate} \\
Note that \texttt{(software} has been treated as a whole term.\\ 

\hline
\end{tabular}
\end{table}

The second bug is the HTTP failure response in the scenario of using a wildcard (*) with less than three characters. In fact, if this scenario involves a single-word term only (e.g., \texttt{en*}), IEEE Xplore can give a user-friendly hint about requiring ``at least 3 valid characters for each keyword''. However, when this scenario exists in a multi-word search term or in a multi-term search string, users will receive a failure response of ``400 Bad Request'' instead of any meaningful message. Even though this case may also be considered as a string bug (i.e.~the user's mistake), the meaningless response can be, and should be, avoided in IEEE Xplore. The simplest workflow of reproducing this bug is shown below.

\begin{table}[!ht]\footnotesize
\renewcommand{\arraystretch}{1.5}
%\caption{An Example of a Table}
%\label{table_example}
\centering
\begin{tabular}{|p{0.95\linewidth}|}
\hline
\textbf{Bug \labelbugs{IEEE2}. Steps to reproduce HTTP error 400 in IEEE Xplore:} (Last test on July 27, 2022)

\begin{enumerate}[leftmargin=4.5ex,after=\vspace{-\baselineskip}]
\item Open IEEE Xplore's homepage (\url{https://ieeexplore.ieee.org/}).
\item Input \texttt{software AND en*} in the default Search field.
\item Click the Search button, and it returns ``Http failure response for https://ieeexplore.ieee.org/rest/search: 400 Bad Request''. 
\item Input \texttt{\textquotedbl software en*\textquotedbl} in the default search field.
\item Click the Search button again, and it still returns ``Http failure response for https://ieeexplore.ieee.org/rest/search: 400 Bad Request''. 
\end{enumerate} \\
\hline
\end{tabular}
\end{table}

\subsection{The Bugs of ScienceDirect}
%\label{subsubsec:bugScienceDirect}
%\vspace{0.5\baselineskip} 
%\noindent
%\textit{5) The Bugs of ScienceDirect}.
ScienceDirect is able, to some extent, to automatically balance the unpaired quotes and parentheses in search strings. In particular, the automatic balancing of unpaired parentheses seems to supplement the missing one at an outermost position. During the verification of this mechanism, we identified a bug related to outermost parentheses in ScienceDirect, i.e.~enclosing a relatively complex search string within a pair of parenthesis will deliver significantly less results than using the original string. It should be noted that this bug is different from the outermost parentheses bug in ACM DL (cf.~Bug \refbugs{ACM2}), and the redundant parentheses have nothing to do with any syntax error. We demonstrate this bug via the steps in Bug \refbugs{SD1}.

\begin{table}[!ht]\footnotesize
\renewcommand{\arraystretch}{1.5}
%\caption{An Example of a Table}
%\label{table_example}
\centering
\begin{tabular}{|p{0.95\linewidth}|}
\hline
\textbf{Bug \labelbugs{SD1}. Steps to reproduce the bug of outermost parentheses interpretation in ScienceDirect:} (Last test on July 28, 2022)

\begin{enumerate}[leftmargin=4.5ex,after=\vspace{-1.3\baselineskip}]
\item Open ScienceDirect's homepage (\url{https://www.sciencedirect.com/}).
\item Input \texttt{software OR engineering AND empirical} in the Keywords field.
\item Click the Search button, and it returns ``3,686,632 results''. 
\item Go back to ScienceDirect's homepage, and input \texttt{(software OR engineering AND empirical)} in the Keywords field.
\item Click the Search button, and it returns ``525,335 results''. 
\end{enumerate} \\
Note that searching for \texttt{software OR (engineering AND empirical)} and \texttt{(software OR engineering) AND empirical} will return ``3,686,634 results'' and ``754,504 results'' respectively.\\

\hline
\end{tabular}
\end{table}

There is no doubt that all the search engines have defined stop words to improve the search efficiency. However, ScienceDirect seems to have inconsistent sets of stop words for different search fields. For example, \texttt{usable} and \texttt{usability} are not searchable in most field-specific cases, whereas they are not in the stop word list \cite{ScienceDirect_2021_stopwords} and they are indeed searchable when searching all parts of the articles. The concise reproduction steps are listed in Bug \refbugs{SD2}. Despite that multiple studies (e.g., \cite{Lima_Salgado_2014}) have employed \texttt{usability} as a search term, a bigger concern is about if ScienceDirect maintains an out-of-date list of more unknown stop words.

\begin{table}[!ht]\footnotesize
\renewcommand{\arraystretch}{1.5}
%\caption{An Example of a Table}
%\label{table_example}
\centering
\begin{tabular}{|p{0.95\linewidth}|}
\hline
\textbf{Bug \labelbugs{SD2}. Steps to reproduce the bug of stop words in ScienceDirect:} (Last test on July 28, 2022)

%\begin{enumerate}[leftmargin=4.5ex,topsep=0.5ex,itemsep=0.5ex,after=\vspace{-1.2\baselineskip}]
\begin{enumerate}[leftmargin=4.5ex,after=\vspace{-1.3\baselineskip}]
\item Open the interface of ScienceDirect's Advanced Search (\url{https://www.sciencedirect.com/search}).
\item Input \texttt{usability} in the top field.
\item Click the Search button, and it returns ``111 results''. 
\item Go back to ScienceDirect's Advanced Search, and input \texttt{usability} in the Title field.
\item Click another field to make the Title field inactive, then the bottom of Title field shows a tip message ``Check for typos or errors'', and the top-left panel shows a warning message ``This search will not return any results.''
\item Click the Search button, nothing happens, or it returns ``No results found.'' 
\end{enumerate} \\
Note that \texttt{usability} is treated as a valid search term in the top field, while as a stop word in the other fields.\\

\hline
\end{tabular}
\end{table}

\subsection{The Bug of Scopus}
%\vspace{0.5\baselineskip} 
%\noindent
%\textit{6) The Bug of Scopus}.
It is known that curly quotes (``~'') may not be interpreted properly as quotation marks in search strings, unless search engines specifically accept them. In Scopus, surprisingly, curly quotes will be treated differently via different search entries: Its advanced search accepts curly quotes while the basic search does not. This inconsistency in a single search engine should be filed as a bug that can be reproduced through the steps in Bug \refbugs{Scopus}.

\begin{table}[!ht]\footnotesize
\renewcommand{\arraystretch}{1.5}
%\caption{An Example of a Table}
%\label{table_example}
\centering
\begin{tabular}{|p{0.95\linewidth}|}
\hline
\textbf{Bug \labelbugs{Scopus}. Steps to reproduce the bug of curly quote interpretation in Scopus:} (Last test on July 24, 2022)

%\begin{enumerate}[leftmargin=4.5ex,topsep=0.5ex,itemsep=0.5ex,after=\vspace{-1.2\baselineskip}]
\begin{enumerate}[leftmargin=4.5ex,after=\vspace{-1.3\baselineskip}]
\item Open Scopus's basic (default) Documents search interface (\url{https://www.scopus.com/search/form.uri?display=basic#basic}).
\item Input \texttt{``software engineering''} in the Search Documents field.
\end{enumerate} \\
Note that the search term \texttt{software engineering} is enclosed within a pair of curly quotes.\\
%\begin{enumerate}[leftmargin=4.5ex,topsep=0.5ex,itemsep=0.5ex,after=\vspace{-1.2\baselineskip},before=\vspace{-0.9\baselineskip}]
\begin{enumerate}[leftmargin=4.5ex,after=\vspace{-1.3\baselineskip},before=\vspace{-1\baselineskip}]
 \setcounter{enumi}{2}
\item Click the Search button, and it returns some amount of hits with TITLE-ABS-KEY(``software AND engineering'').
\item Click the Go Back button of the web browser, and go back to the basic search interface. 
\item Click the Advanced Document Search link, and go to the advanced search interface. 
\end{enumerate} \\
Note that the Enter Query String field shows TITLE-ABS-KEY(``software engineering'') by default.\\ 
%\begin{enumerate}[leftmargin=4.5ex,topsep=0.5ex,itemsep=0.5ex,after=\vspace{-1.2\baselineskip},before=\vspace{-0.9\baselineskip}]
\begin{enumerate}[leftmargin=4.5ex,after=\vspace{-1.3\baselineskip},before=\vspace{-1\baselineskip}]
 \setcounter{enumi}{5}
\item Click the Search button, and it returns much less hits with TITLE-ABS-KEY(\textquotedbl software engineering\textquotedbl). 
\end{enumerate} \\
Note that the quotation mark format has been changed automatically from curly quotes into straight quotes.\\
\hline
\end{tabular}
\end{table}

\subsection{The Bugs of SpringerLink}
%\label{subsubsec:bugSpringer}
%\vspace{0.5\baselineskip} 
%\noindent
%\textit{7) The Bugs of SpringerLink}.
As mentioned previously, title search is the only field-specific search option in SpringerLink. Although we do not treat any functional limit as a bug, it is worth reminding practitioners of the misleading title search functionality of SpringerLink: Firstly, its title search does not work with Boolean operators, unless one of the search terms is enclosed within a pair of quotation marks (\textquotedbl ~\textquotedbl). Secondly, most of the title search results do not match the original search string. The steps to reproduce this bug are specified in Bug \refbugs{SL1}. 

\begin{table}[!ht]\footnotesize
\renewcommand{\arraystretch}{1.5}
%\caption{An Example of a Table}
%\label{table_example}
\centering
\begin{tabular}{|p{0.95\linewidth}|}
\hline
\textbf{Bug \labelbugs{SL1}. Steps to reproduce the bug of title search in SpringerLink:} (Last test on July 22, 2022)

%\begin{enumerate}[leftmargin=4.5ex,topsep=0.5ex,itemsep=0.5ex,after=\vspace{-1.2\baselineskip}]
\begin{enumerate}[leftmargin=4.5ex,after=\vspace{-1.3\baselineskip}]
\item Open the interface of SpringerLink's Advanced Search (\url{https://link.springer.com/advanced-search}). 
\item Go to the field ``where the \textbf{title} contains'' and input the search string \texttt{software AND engineering}.
\item Click the Search button, and it returns ``0 Result(s)''.
\item Go back and remove \texttt{AND} from the search string.
\item Click the Search button, and it returns ``2,239 Result(s)''.
\end{enumerate} \\
Note that \texttt{software engineering} has been treated as a single search term.\\
%\begin{enumerate}[leftmargin=4.5ex,topsep=0.5ex,itemsep=0.5ex,after=\vspace{-1.2\baselineskip},before=\vspace{-0.9\baselineskip}]
\begin{enumerate}[leftmargin=4.5ex,after=\vspace{-1.3\baselineskip},before=\vspace{-1.1\baselineskip}]
 \setcounter{enumi}{5}
\item Go back and change the search string into \texttt{software AND \textquotedbl engineering\textquotedbl}.
\item Click the Search button, and it returns ``25,801 Result(s)''.
\end{enumerate} \\
Note that most results do not contain the both search terms in the publication titles.\\
\hline
\end{tabular}
\end{table}

Another bug is related to SpringerLink's vague capacity for handling extremely long search strings (the threshold seems to be about 1800 characters according to our tests). Although SpringerLink does not seem to have any constraint on string length, refining the default search results from long strings will be rejected with an HTTP error 431 ``Request Header Fields Too Large''. Since this error may have various reasons at the server side \cite{Okta_2021}, we are not able to reveal the root cause of this bug, but only giving its reproduction steps as shown in Bug \refbugs{SL2}.

\begin{table}[!ht]\footnotesize
\renewcommand{\arraystretch}{1.5}
%\caption{An Example of a Table}
%\label{table_example}
\centering
\begin{tabular}{|p{0.95\linewidth}|}
\hline
\textbf{Bug \labelbugs{SL2}. Steps to reproduce HTTP error 431 in SpringerLink:} (Last test on July 21, 2022)

%\begin{enumerate}[leftmargin=4.5ex,topsep=0.5ex,itemsep=0.5ex,after=\vspace{-\baselineskip}]
\begin{enumerate}[leftmargin=4.5ex,after=\vspace{-\baselineskip}]
\item Open SpringerLink's homepage (\url{https://link.springer.com/}). 
\item Input a lengthy search string (over about 1800 characters) in, or directly copy the search string from \cite{Ferreira_2017} to, the Search field.
\item Click the Search button, and the search results should be able to show up.
\item Refine the search results by clicking an item (e.g., ``Article'') within the left side Refine Your Search panel.
\item The page returns ``Bad Message 431'' and ``reason: Request Header Fields Too Large''.
\end{enumerate} \\
\hline
\end{tabular}
\end{table}

\subsection{The Bug of Wiley Online Library}
%\vspace{0.5\baselineskip} 
%\noindent
%\textit{8) The Bug of Wiley Online Library}.
Although the repeatability of filter settings is not measured in this research, we have carefully tried filtering the publishing venues for those studies who employed venue-specific automated search. During those trials, we observed a bug in Wiley OL: The pre-set filter of publishing venue will not be applied to the search results. It should be noted that the dysfunctional venue filtering can be misleading or at least a waste of time for users, even though it can be fixed by selecting filters again on the results page. The steps to reproduce this bug are specified in Bug \refbugs{Wiley}.

\begin{table}[!ht]\footnotesize
\renewcommand{\arraystretch}{1.5}
%\caption{An Example of a Table}
%\label{table_example}
\centering
\begin{tabular}{|p{0.95\linewidth}|}
\hline
\textbf{Bug \labelbugs{Wiley}. Steps to reproduce the bug of venue filtering in Wiley OL:} (Last test on July 28, 2022)

%\begin{enumerate}[leftmargin=4.5ex,topsep=0.5ex,itemsep=0.5ex,after=\vspace{-1.2\baselineskip}]
\begin{enumerate}[leftmargin=4.5ex,after=\vspace{-1.3\baselineskip}]
\item Open the interface of Wiley OL's Advanced Search (\url{https://onlinelibrary.wiley.com/search/advanced}). 
\item Input \texttt{software} in the top search field.
\item Input \texttt{Software: Practice and Experience} in the venue field under ``Published in''.
\item Click the Search button, and then scroll down to the bottom of the Filters panel in the search results page.
\end{enumerate} \\
Note that all the publishing venues are listed, which mean that the pre-set venue filter is not applied to the results.\\

\hline
\end{tabular}
\end{table}

\section{Recommendations to the Developers of Academic Search Engines}
\label{sec:suggestions}
Since search engines would never stop evolving, a possible concern is about the consistency issues in the user interaction due to the changing conditions (i.e.~the search engines' interfaces and/or functionalities). We believe that this concern implies a pessimistic assumption about the influence of search engine evolution. It is true that some engines' functional or interface changes have made previous user interactions non-replicable. However, the evolution of some others has brought more user-friendly features. For instance, the generic string in \cite{Bianchi_2015} was previously required to be customised for ACM DL, while it can now be used directly without any modification. Therefore, we decided to briefly summarise the past changes of academic search engines, and then use them to strengthen our evidence-based recommendations about the future evolution directions.% of search engine development. 
  
\subsection{Publications Have a Memory: The Past Changes}
\label{subsec:evolutionHistory}
Although we do not physically track the historical changes in the existing digital libraries and indexing platforms, we have observed various differences between the previously published details and our own user experience. These differences can then be used to reveal the evolution of relevant search engines, to some extent. From the perspective of enhancing replicability (that may indicate the improvement of backward compatibility of a single engine and/or the improvement of compatibility among multiple engines), we distinguish the past changes between evolution and devolution of search engines.

\subsubsection{Evolution}
\label{subsubsec:improvements}
Firstly, we see a trend in unifying the interface of advanced search. Although there is still inconsistent functionality (e.g., ACM DL supports command search, while Wiley OL does not), multiple digital libraries seem to have employed the same and new interface design, such as ACM DL (\url{https://dl.acm.org/search/advanced}), INFORMS PubsOnLine (\url{https://pubsonline.informs.org/search/advanced}), Wiley OL (\url{https://onlinelibrary.wiley.com/search/advanced}), and World Scientific (\url{https://www.worldscientific.com/search/advanced}). 

Secondly, there is a clear trend in unifying the Boolean operators for combining search terms. The lowercase format used in ACM DL \cite{Haselberger_2016} and the angle-bracketed lowercase version used in IEEE Xplore \cite{Budgen_2011} have all been changed into the capital-case Boolean operators that are the standard format in almost all the search engines.

Thirdly, there also seems a trend in unifying zonal settings. For instance, the keyword search was not an option in ACM DL and in Wiley OL at least before 2015 \cite{Yang_Liang_2016}. Furthermore, ACM DL's full text search was unavailable at least until the year of 2011 \cite{Rinkevics_2013}. Although the search field of Review in ACM DL \cite{Lucas_Molina_2009} is removed, not many secondary studies have employed this field in their search activities.

At last, some search engine constraints documented in the previous publications are relaxed.
For example, nowadays users are able to input a long and complex string in any search field of ACM DL, which was not supported before according to \cite{Chen_Ali_2009}. Similarly, IEEE Xplore did not support lengthy search strings even in the year of 2015 \cite{Soomro_2016}, while it works well now. Particularly, in this research, we observed that IEEE Xplore's maximum limit of wildcards (*) was increased from six to seven in the past year.

\subsubsection{Devolution}
\label{subsubsec:degeneration}
It is understandable that none of the publishers and indexers will claim their changes to be devolution. Unfortunately, the changes of some search engines have introduced more troubles (or at least constraints) to the current automated searches. One of the typical issues is the removal of popular search fields. For example, the keyword search and abstract search used to be functional in SpringerLink (e.g., \cite{Kusumo_2012}), while they are both unavailable at the time of this research. The available title search is misleading: Firstly, SpringerLink's title search does not work with Boolean operators, unless one of the search terms is enclosed within a pair of quotation marks (\textquotedbl ~\textquotedbl). Secondly, most of the title search results may not match the search string. Thus, we have to claim that SpringerLink does not (well) support metadata search.   

Another typical issue is the increased constraints of the searching features. For instance, ScienceDirect used to not only support long strings (e.g., \cite{Khan_Niazi_2011_IST}), but also to allow wildcards (*) (e.g., \cite{Garcia_2014}) and more than eight Boolean operators (e.g., \cite{Santos_2018}) to be used in search strings. Nevertheless, these features have all been limited, according to our tests.

%At the time of this research, SpringerLink supports either default search or title search. In particular, according to our tests, the default search seems to be equivalent to the all-field search. Since some early studies claim to have conducted keyword and/or abstract searches in SpringerLink (e.g., \cite{Kusumo_2012}), there might also be functional degradation of SpringerLink's search engine. However, considering that a study \cite{Maplesden_2015} performed on 18 December 2013 specifically confirms the current user experience, it is also possible that the later work misreported their automatic search implementations (e.g., \cite{Garcia_2016,Mariani_2017}). 

\subsection{Recommendations for the Future Evolution}
\label{subsec:sourceSuggestions}
After summarising the aforementioned observations and our user experience in the automated search tests, we come up with two groups of recommendations for evolving search engines in the future. The first group is towards researcher-friendly evolution, including:

\begin{itemize}%[leftmargin=*]
\item \textbf{Relaxing the current constraints}. As mentioned previously, the search engine diversity and constraints frequently require modifying (e.g., shortening or adapting) well-designed search strings, while it is not a trivial task to mix major terms and synonyms properly \cite{Imtiaz_2013}. Considering that systematic reviews pervasively need long and sophisticated search strings, it had better relax the search engine constraints of string length and Boolean operations. In addition, offering the feature of command line search would be particularly useful for manipulating and running long and sophisticated search strings.

\item \textbf{Syntax validation of search strings}. Human beings make mistakes. As demonstrated by Bug \refbugs{Scopus}, researchers can incidentally introduce various logic and stylistic errors especially to long and complex search strings. Therefore, offering syntax validation will significantly facilitate researchers to use correct strings to conduct automated search. 
In fact, our tests have found that Scopus and WoS are equipped with relatively well developed mechanisms of syntax validation.

\item \textbf{Unified and standard syntax and syntax rules}. There is no doubt that diverse syntax and rules of different search engines (e.g., DBLP and Engineering Village) will inevitably increase the workloads of building search strings, even if researchers are able to equivalently translate one into another. Therefore, we urge unified and standard syntax and syntax rules to improve the user experience across multiple search engines. In fact, as mentioned in Section \ref{subsubsec:improvements}, most search engines have been evolving along this direction. 

\item \textbf{Unified and standard front-end interfaces}. Compared with the syntax inconsistency, the diverse search engine interfaces and engine features seem to bother researchers more frequently and cause more issues (e.g., search string adaptation) in automated search. Thus, unifying interfaces will at least ease the user operations when a user needs to switch from one search engine to another.  

\item \textbf{Fully featured APIs}. Unlike ad hoc reviews, systematic reviews generally employ automated search for and access massive number of papers.  Ideally, programmably searching and obtaining the search results will make automated search really automatic. Unfortunately, compared with the promising progress for automating paper selection \cite{Shakeel_2019}, the current automated searches still largely rely on manual operations via search engines' user interfaces. Note that crawling web pages is widely prohibited by digital libraries, while not all the search engines offer APIs. Among the limited available APIs, most of them are not fully featured (e.g., only metadata search is supported). %, not to mention any version control mechanism (e.g., the temporal consistency issue in the search results via APIs \cite{Kruger_2020}). %researchers are professional users of digital libraries and indexing platforms.
\end{itemize}

By facilitating operations and by reducing human mistakes, the researcher-friendly evolution will indirectly help enhance the replicability of user interactions. In contrast, the second group of recommendations is specifically towards replicability-oriented evolution, including:
\begin{itemize}%[leftmargin=*]
\item \textbf{Backward compatible updates}. As mentioned previously, replicating user interactions relies not only on the same input, but also on the same/similar environments or conditions.  If a search engine is updated without remaining compatible with its earlier functional features, it will be difficult to expect previous practices to be replicable on the same search engine. For example, according to the explanation \cite{Elsevier_2020}, the increased search constraints of ScienceDirect may be a compromise to improve its engine performance, which however has significantly sacrificed the replicability of user interaction in many existing studies \cite{Li_2021}. 

\item \textbf{Unified and standard back-end functional features}. As mentioned above, the inconsistent search engine features and interfaces not only complicate the search process but also incur replicability problems. For example, search string adaptation has been identified as a major factor that threatens search replications \cite{Kruger_2020}. Moreover, the engine features will have cascaded influences on the design and the usage of APIs. Considering that academic publications are well structured data with standard metadata, it is feasible to unify the functional features of academic search engines, to better support the unification of interfaces.
\end{itemize}
%"and not the other way around.

Overall, our recommendations about unification and standardisation further require extensive collaboration among those publishers and indexers. Although there has been spontaneous consensus among them, the community may need an independent committee to help them cooperate (and also possibly negotiate) with each other more efficiently.

%%%%%%%%%%%%%%%%%%%%%%%%%%%%%%%%%%%%%%%%%%%%%%%%%

\section{Discussion and Conclusions}% and Future Work}
\label{sec:conclusions}

\subsection{Discussion}
%\textbf{\textit{Discussion:}}

It is well recognised that the evaluation of search engines (especially with respect to digital libraries) is both useful and challenging, because it has various criteria and standards for verifying different attributes, aspects, and components from different perspectives \cite{Li_Liu_2019}. We acknowledge that our work only conducts user interaction testing and reflects the effectiveness part of search engine usability, without measuring the other evaluation dimensions.

However, given our research resource (manpower) limit, it is impractical to cover a wide range of evaluation dimensions for such extensive testing work. On the other hand, it is common for researchers to intentionally exclude measures from search engine evaluation in practice. For example, \textit{recall} and \textit{precision} have frequently been omitted in large scale testing \cite{Vaughan_2004}, because exhaustive judgements for document relevance against every single request are infeasible with large-size literature collections. Therefore, the theoretical evaluation models and frameworks (see Section \ref{sec:RelatedWork}) may need a divide-and-conquer fashion to guide practices.

Following the divide-and-conquer fashion, we have chosen to investigate the most critical attribute (i.e.~effectiveness) of search engines' usability. As one of the three main themes of search engine evaluation (the other two are the value of collection and the system performance) \cite{Xie_2008}, usability measures how usable and how easy-to-use a search engine is. Before proving a search engine to be effectively usable, it will be meaningless to discuss its easiness of usage.

\subsection{Conclusions and Future Work}
%\noindent
%\textbf{\textit{Conclusions:}}

Our study leads to two conclusions. First, there remains a considerable gap between the search-needs of researchers and the usability of academic search engines. It is not clear whether search-engine developers are aware of this continuing gap. An early tertiary study has already highlighted academic search engines' poor support to systematic reviews \cite{Brereton_2007}. Our study provides a recent complement to the tertiary review, and shows that problems with search engines remain.

Second, the \textit{evaluation}, by academics, of academic search engines has not kept pace with the \textit{development}, by search-engine providers, of those search engines. This gap between evaluation and development makes it harder to properly understand the gap between the search-needs of researchers and search-features of the search engines.

% has significantly fallen behind the engine development and applications in the scientific community, which in turn deepens the understanding gap between the engine owners and the researchers.  

Compared with general search engines, the academic engines (should) have unique features that are particularly designed for the specific requirements of its (academic) users, e.g., that the results of a search are exhaustive, or that the results are ranked in a way more relevant and reliable for research, rather than in a way that is more convenient or quicker to process. In other words, academics have a higher performance requirement of their search engines, compared to public usage of search engines, and search-engine providers should have an appreciation for these specific performance requirements.

% intended for and by their intended users. For example, the scientific usage of search engines tends to employ well-prepared and sophisticated search strings and expects highly relevant hits. In contrast, %although precision is one of the standard evaluation criteria for information retrieval in general, 
% the public usage of search engines normally chooses short and vague search strings and thus may be tolerant of flaws in the search results.

%\subsection{Future Work}
%\noindent
%\textbf{\textit{Future Work:}}

Driven by the conclusions, we argue the needs of dedicated efforts on evaluating academic search engines.
Therefore, our next-step work will be distilling a systematic yet concise suite of test cases from this random-sample-based regression testing. Then, we plan not only to share the test case suite with colleague researchers, but also to contact the relevant search engine providers and help them address the diverse needs from the scientific community of which they are also part.   

%\textcolor{Maroon}{Conclusions:   Future work: Automated testing scripts/tools will be developed based on a set of well-defined test cases.}

%The automation of testing will have to depend on the APIs offered by the digital libraries, while
\begin{acks}
This research is supported in part by startup funding from Queen's University Belfast. We also thank reviewers, in advance, for their reviews.
% the project D8203EEC with sub-analysis 3057248.?????
\end{acks}

\bibliographystyle{ACM-Reference-Format}
\bibliography{sample-base,ATESTref}


\begin{thebibliography}{40}

%%% the Web address.

\ifx \showCODEN    \undefined \def \showCODEN     #1{\unskip}     \fi
\ifx \showDOI      \undefined \def \showDOI       #1{#1}\fi
\ifx \showISBNx    \undefined \def \showISBNx     #1{\unskip}     \fi
\ifx \showISBNxiii \undefined \def \showISBNxiii  #1{\unskip}     \fi
\ifx \showISSN     \undefined \def \showISSN      #1{\unskip}     \fi
\ifx \showLCCN     \undefined \def \showLCCN      #1{\unskip}     \fi
\ifx \shownote     \undefined \def \shownote      #1{#1}          \fi
\ifx \showarticletitle \undefined \def \showarticletitle #1{#1}   \fi
\ifx \showURL      \undefined \def \showURL       {\relax}        \fi
% invisible to TeX
\providecommand\bibfield[2]{#2}
\providecommand\bibinfo[2]{#2}
\providecommand\natexlab[1]{#1}
\providecommand\showeprint[2][]{arXiv:#2}

\bibitem[Bailey et~al\mbox{.}(2007)]%
        {Bailey_Zhang_2007}
\bibfield{author}{\bibinfo{person}{John Bailey}, \bibinfo{person}{Cheng Zhang},
  \bibinfo{person}{David Budgen}, \bibinfo{person}{Mark Turner}, {and}
  \bibinfo{person}{Stuart Charters}.} \bibinfo{year}{2007}\natexlab{}.
\newblock \showarticletitle{Search Engine Overlaps: Do they agree or
  disagree?}. In \bibinfo{booktitle}{\emph{Proceedings of the 2nd International
  Workshop on Realising Evidence-Based Software Engineering (REBSE 2007)}}.
  \bibinfo{publisher}{IEEE Computer Society}, \bibinfo{address}{Minneapolis,
  MN, USA}, \bibinfo{pages}{1--6}.
\newblock


\bibitem[Bianchi et~al\mbox{.}(2015)]%
        {Bianchi_2015}
\bibfield{author}{\bibinfo{person}{Thiago Bianchi},
  \bibinfo{person}{Daniel~Soares Santos}, {and} \bibinfo{person}{Katia~Romero
  Felizardo}.} \bibinfo{year}{2015}\natexlab{}.
\newblock \showarticletitle{Quality Attributes of Systems-of-Systems: A
  Systematic Literature Review}. In \bibinfo{booktitle}{\emph{Proceedings of
  the 3rd IEEE/ACM International Workshop on Software Engineering for
  Systems-of-Systems (SESoS 2015)}}. \bibinfo{publisher}{IEEE Press},
  \bibinfo{address}{Florence, Italy}, \bibinfo{pages}{23--30}.
\newblock


\bibitem[Brereton et~al\mbox{.}(2007)]%
        {Brereton_2007}
\bibfield{author}{\bibinfo{person}{Pearl Brereton}, \bibinfo{person}{Barbara~A.
  Kitchenham}, \bibinfo{person}{David Budgen}, \bibinfo{person}{Mark Turner},
  {and} \bibinfo{person}{Mohamed Khalil}.} \bibinfo{year}{2007}\natexlab{}.
\newblock \showarticletitle{Lessons from applying the systematic literature
  review process within the software engineering domain}.
\newblock \bibinfo{journal}{\emph{Journal of Systems and Software}}
  \bibinfo{volume}{80}, \bibinfo{number}{4} (\bibinfo{date}{April}
  \bibinfo{year}{2007}), \bibinfo{pages}{571--583}.
\newblock


\bibitem[Buchanan and Salako(2009)]%
        {Buchanan_2009}
\bibfield{author}{\bibinfo{person}{Steven Buchanan} {and}
  \bibinfo{person}{Adeola Salako}.} \bibinfo{year}{2009}\natexlab{}.
\newblock \showarticletitle{Evaluating the usability and usefulness of a
  digital library}.
\newblock \bibinfo{journal}{\emph{Library Review}} \bibinfo{volume}{58},
  \bibinfo{number}{9} (\bibinfo{date}{October} \bibinfo{year}{2009}),
  \bibinfo{pages}{638--651}.
\newblock


\bibitem[Budgen et~al\mbox{.}(2011)]%
        {Budgen_2011}
\bibfield{author}{\bibinfo{person}{D. Budgen}, \bibinfo{person}{A.~J. Burn},
  \bibinfo{person}{O.~P. Brereton}, \bibinfo{person}{B.~A. Kitchenham}, {and}
  \bibinfo{person}{R. Pretorius}.} \bibinfo{year}{2011}\natexlab{}.
\newblock \showarticletitle{Empirical evidence about the {UML}: A systematic
  literature review}.
\newblock \bibinfo{journal}{\emph{Software: Practice and Experience}}
  \bibinfo{volume}{41}, \bibinfo{number}{4} (\bibinfo{date}{April}
  \bibinfo{year}{2011}), \bibinfo{pages}{363--392}.
\newblock


\bibitem[Chen et~al\mbox{.}(2009)]%
        {Chen_Ali_2009}
\bibfield{author}{\bibinfo{person}{Lianping Chen},
  \bibinfo{person}{Muhammad~Ali Babar}, {and} \bibinfo{person}{Ciaran Cawley}.}
  \bibinfo{year}{2009}\natexlab{}.
\newblock \showarticletitle{A Status Report on the Evaluation of Variability
  Management Approaches}. In \bibinfo{booktitle}{\emph{Proceedings of the 13th
  International Conference on Evaluation and Assessment in Software Engineering
  (EASE 2009)}}. \bibinfo{publisher}{BCS Learning \& Development Ltd.},
  \bibinfo{address}{Durham University, UK}, \bibinfo{pages}{1--10}.
\newblock


\bibitem[Chen et~al\mbox{.}(2010)]%
        {Chen_Babar_2010}
\bibfield{author}{\bibinfo{person}{Lianipng Chen},
  \bibinfo{person}{Muhammad~Ali Babar}, {and} \bibinfo{person}{He Zhang}.}
  \bibinfo{year}{2010}\natexlab{}.
\newblock \showarticletitle{Towards an evidence-based understanding of
  electronic data sources}. In \bibinfo{booktitle}{\emph{Proceedings of the
  14th international conference on Evaluation and Assessment in Software
  Engineering (EASE 2010)}}. \bibinfo{publisher}{BCS}, \bibinfo{address}{Keele
  University, UK}, \bibinfo{pages}{135--138}.
\newblock


\bibitem[de~Lima~Salgado and Freire(2014)]%
        {Lima_Salgado_2014}
\bibfield{author}{\bibinfo{person}{Andr{\'e} de Lima~Salgado} {and}
  \bibinfo{person}{Andr{\'e}~Pimenta Freire}.} \bibinfo{year}{2014}\natexlab{}.
\newblock \showarticletitle{Heuristic Evaluation of Mobile Usability: A Mapping
  Study}.
\newblock In \bibinfo{booktitle}{\emph{Human-Computer Interaction. Applications
  and Services. HCI 2014}}, \bibfield{editor}{\bibinfo{person}{Masaaki Kurosu}}
  (Ed.). \bibinfo{series}{Lecture Notes in Computer Science},
  Vol.~\bibinfo{volume}{8512}. \bibinfo{publisher}{Springer, Cham},
  \bibinfo{address}{Heraklion, Crete, Greece}, \bibinfo{pages}{178--188}.
\newblock


\bibitem[Dyb\r{a} et~al\mbox{.}(2007)]%
        {Dyba_2007}
\bibfield{author}{\bibinfo{person}{Tore Dyb\r{a}}, \bibinfo{person}{Torgeir
  Dings\o{}yr}, {and} \bibinfo{person}{Geir~K. Hanssen}.}
  \bibinfo{year}{2007}\natexlab{}.
\newblock \showarticletitle{Applying Systematic Reviews to Diverse Study Types:
  An Experience Report}. In \bibinfo{booktitle}{\emph{Proceedings of the 1st
  International Symposium on Empirical Software Engineering and Measurement
  (ESEM 2007)}}. \bibinfo{publisher}{IEEE Computer Society},
  \bibinfo{address}{Madrid, Spain}, \bibinfo{pages}{225--234}.
\newblock


\bibitem[Elsevier(2020)]%
        {Elsevier_2020}
\bibfield{author}{\bibinfo{person}{Elsevier}.} \bibinfo{year}{2020}\natexlab{}.
\newblock \bibinfo{title}{New ScienceDirect advanced search delivers fast and
  powerful results}.
\newblock
  \bibinfo{howpublished}{\url{https://www.elsevier.com/librarians/article-news/new-sciencedirect-advanced-search-delivers-fast-and-powerful-results}}.
\newblock


\bibitem[Ferreira et~al\mbox{.}(2017)]%
        {Ferreira_2017}
\bibfield{author}{\bibinfo{person}{Thiago~Nascimento Ferreira},
  \bibinfo{person}{Silvia~Regina Vergilio}, {and}
  \bibinfo{person}{Jerffeson~Teixeira de Souza}.}
  \bibinfo{year}{2017}\natexlab{}.
\newblock \showarticletitle{Incorporating user preferences in search-based
  software engineering: A systematic mapping study}.
\newblock \bibinfo{journal}{\emph{Information and Software Technology}}
  \bibinfo{volume}{90} (\bibinfo{date}{October} \bibinfo{year}{2017}),
  \bibinfo{pages}{55--69}.
\newblock


\bibitem[Garcia et~al\mbox{.}(2014)]%
        {Garcia_2014}
\bibfield{author}{\bibinfo{person}{Cecilia Garcia}, \bibinfo{person}{Abraham
  D{\'a}vila}, {and} \bibinfo{person}{Marcelo Pessoa}.}
  \bibinfo{year}{2014}\natexlab{}.
\newblock \showarticletitle{Test Process Models: Systematic Literature Review}.
  In \bibinfo{booktitle}{\emph{Proceedings of the 14th International Conference
  on Software Process Improvement and Capability Determination (SPICE 2014)}}.
  \bibinfo{publisher}{Springer, Cham}, \bibinfo{address}{Vilnius, Lithuania},
  \bibinfo{pages}{84--93}.
\newblock


\bibitem[Hamilton(2017)]%
        {Hamilton_2017}
\bibfield{author}{\bibinfo{person}{Brian Hamilton}.}
  \bibinfo{year}{2017}\natexlab{}.
\newblock \bibinfo{title}{The Importance of Steps to Reproduce a Bug}.
\newblock
  \bibinfo{howpublished}{\url{https://blog.testlodge.com/the-importance-of-steps-to-reproduce-a-bug/}}.
\newblock


\bibitem[Haselberger(2016)]%
        {Haselberger_2016}
\bibfield{author}{\bibinfo{person}{David Haselberger}.}
  \bibinfo{year}{2016}\natexlab{}.
\newblock \showarticletitle{A literature-based framework of performance-related
  leadership interactions in {ICT} project teams}.
\newblock \bibinfo{journal}{\emph{Information and Software Technology}}
  \bibinfo{volume}{70} (\bibinfo{date}{February} \bibinfo{year}{2016}),
  \bibinfo{pages}{1--17}.
\newblock


\bibitem[Imtiaz et~al\mbox{.}(2013)]%
        {Imtiaz_2013}
\bibfield{author}{\bibinfo{person}{Salma Imtiaz}, \bibinfo{person}{Muneera
  Bano}, \bibinfo{person}{Naveed Ikram}, {and} \bibinfo{person}{Mahmood
  Niazi}.} \bibinfo{year}{2013}\natexlab{}.
\newblock \showarticletitle{A Tertiary Study: Experiences of Conducting
  Systematic Literature Reviews in Software Engineering}. In
  \bibinfo{booktitle}{\emph{Proceedings of the 17th International Conference on
  Evaluation and Assessment in Software Engineering (EASE 2013)}}.
  \bibinfo{publisher}{ACM Press}, \bibinfo{address}{Porto de Galinhas, Brazil},
  \bibinfo{pages}{177--182}.
\newblock


\bibitem[Jeng(2005)]%
        {Jeng_2005}
\bibfield{author}{\bibinfo{person}{Judy Jeng}.}
  \bibinfo{year}{2005}\natexlab{}.
\newblock \showarticletitle{What Is Usability in the Context of the Digital
  Library and How Can It Be Measured}.
\newblock \bibinfo{journal}{\emph{Information Technology and Libraries}}
  \bibinfo{volume}{24}, \bibinfo{number}{2} (\bibinfo{date}{June}
  \bibinfo{year}{2005}), \bibinfo{pages}{47--56}.
\newblock


\bibitem[Kelly(2014)]%
        {Kelly_2014}
\bibfield{author}{\bibinfo{person}{Elizabeth~Joan Kelly}.}
  \bibinfo{year}{2014}\natexlab{}.
\newblock \showarticletitle{Assessment of Digitized Library and Archives
  Materials: A Literature Review}.
\newblock \bibinfo{journal}{\emph{Journal of Web Librarianship}}
  \bibinfo{volume}{8}, \bibinfo{number}{4} (\bibinfo{date}{September}
  \bibinfo{year}{2014}), \bibinfo{pages}{384--403}.
\newblock


\bibitem[Khan et~al\mbox{.}(2011)]%
        {Khan_Niazi_2011_IST}
\bibfield{author}{\bibinfo{person}{Siffat~Ullah Khan}, \bibinfo{person}{Mahmood
  Niazi}, {and} \bibinfo{person}{Rashid Ahmad}.}
  \bibinfo{year}{2011}\natexlab{}.
\newblock \showarticletitle{Barriers in the selection of offshore software
  development outsourcing vendors: An exploratory study using a systematic
  literature review}.
\newblock \bibinfo{journal}{\emph{Information and Software Technology}}
  \bibinfo{volume}{53}, \bibinfo{number}{7} (\bibinfo{date}{July}
  \bibinfo{year}{2011}), \bibinfo{pages}{693--706}.
\newblock


\bibitem[Khoo et~al\mbox{.}(2012)]%
        {Khoo_2012}
\bibfield{author}{\bibinfo{person}{Michael Khoo}, \bibinfo{person}{Diana
  Kusunoki}, {and} \bibinfo{person}{Craig MacDonald}.}
  \bibinfo{year}{2012}\natexlab{}.
\newblock \showarticletitle{Finding Problems: When Digital Library Users Act as
  Usability Evaluators}. In \bibinfo{booktitle}{\emph{Proceedings of the 45th
  Hawaii International Conference on System Sciences (HICSS 2012)}}.
  \bibinfo{publisher}{IEEE Computer Society}, \bibinfo{address}{Maui, HI, USA},
  \bibinfo{pages}{1615--1624}.
\newblock


\bibitem[Kitchenham and Charters(2007)]%
        {Kitchenham_Charters_2007}
\bibfield{author}{\bibinfo{person}{Barbara~A. Kitchenham} {and}
  \bibinfo{person}{Stuart Charters}.} \bibinfo{year}{2007}\natexlab{}.
\newblock \bibinfo{booktitle}{\emph{Guidelines for Performing Systematic
  Literature Reviews in Software Engineering}}.
\newblock \bibinfo{type}{Technical Report} EBSE 2007-01.
  \bibinfo{institution}{School of Computer Science and Mathematics, Keele
  University and Durham University}.
\newblock


\bibitem[Kr{\"u}ger et~al\mbox{.}(2020)]%
        {Kruger_2020}
\bibfield{author}{\bibinfo{person}{Jacob Kr{\"u}ger},
  \bibinfo{person}{Christian Lausberger}, \bibinfo{person}{Ivonne von
  Nostitz-Wallwitz}, \bibinfo{person}{Gunter Saake}, {and}
  \bibinfo{person}{Thomas Leich}.} \bibinfo{year}{2020}\natexlab{}.
\newblock \showarticletitle{Search. Review. Repeat? An empirical study of
  threats to replicating {SLR} searches}.
\newblock \bibinfo{journal}{\emph{Empirical Software Engineering}}
  \bibinfo{volume}{25} (\bibinfo{date}{January} \bibinfo{year}{2020}),
  \bibinfo{pages}{627--677}.
\newblock


\bibitem[Kusumo et~al\mbox{.}(2012)]%
        {Kusumo_2012}
\bibfield{author}{\bibinfo{person}{Dana~S. Kusumo}, \bibinfo{person}{Mark
  Staples}, \bibinfo{person}{Liming Zhu}, \bibinfo{person}{He Zhang}, {and}
  \bibinfo{person}{Ross Jeffery}.} \bibinfo{year}{2012}\natexlab{}.
\newblock \showarticletitle{Risks of off-the-shelf-based software acquisition
  and development: A systematic mapping study and a survey}. In
  \bibinfo{booktitle}{\emph{Proceedings of the 16th International Conference on
  Evaluation and Assessment in Software Engineering (EASE 2012)}}.
  \bibinfo{publisher}{IET press}, \bibinfo{address}{Ciudad Real, Spain},
  \bibinfo{pages}{233--242}.
\newblock


\bibitem[Lai et~al\mbox{.}(2014)]%
        {Lai_2014}
\bibfield{author}{\bibinfo{person}{Chin-Feng Lai}, \bibinfo{person}{Po-Sheng
  Chiu}, \bibinfo{person}{Yueh-Min Huang}, \bibinfo{person}{Tzung-Shi Chen},
  {and} \bibinfo{person}{Tien-Chi Huang}.} \bibinfo{year}{2014}\natexlab{}.
\newblock \showarticletitle{An evaluation model for digital libraries' user
  interfaces using fuzzy {AHP}}.
\newblock \bibinfo{journal}{\emph{The Electronic Library}}
  \bibinfo{volume}{32}, \bibinfo{number}{1} (\bibinfo{date}{January}
  \bibinfo{year}{2014}), \bibinfo{pages}{83--95}.
\newblock


\bibitem[Li and Liu(2019)]%
        {Li_Liu_2019}
\bibfield{author}{\bibinfo{person}{Yuelin Li} {and} \bibinfo{person}{Chang
  Liu}.} \bibinfo{year}{2019}\natexlab{}.
\newblock \showarticletitle{Information Resource, Interface, and Tasks as User
  Interaction Components for Digital Library Evaluation}.
\newblock \bibinfo{journal}{\emph{Information Processing and Management}}
  \bibinfo{volume}{56}, \bibinfo{number}{3} (\bibinfo{date}{May}
  \bibinfo{year}{2019}), \bibinfo{pages}{704--720}.
\newblock


\bibitem[Li(2021)]%
        {Li_2021}
\bibfield{author}{\bibinfo{person}{Zheng Li}.} \bibinfo{year}{2021}\natexlab{}.
\newblock \showarticletitle{Stop Building Castles on a Swamp! The Crisis of
  Reproducing Automatic Search in Evidence-based Software Engineering}. In
  \bibinfo{booktitle}{\emph{Proceedings of the 43rd IEEE/ACM International
  Conference on Software Engineering: New Ideas and Emerging Results (ICSE-NIER
  2021)}}. \bibinfo{publisher}{IEEE Press}, \bibinfo{address}{Madrid, Spain},
  \bibinfo{pages}{16--20}.
\newblock


\bibitem[Lucas et~al\mbox{.}(2009)]%
        {Lucas_Molina_2009}
\bibfield{author}{\bibinfo{person}{Francisco~J. Lucas},
  \bibinfo{person}{Fernando Molina}, {and} \bibinfo{person}{Ambrosio Toval}.}
  \bibinfo{year}{2009}\natexlab{}.
\newblock \showarticletitle{A systematic review of UML model consistency
  management}.
\newblock \bibinfo{journal}{\emph{Information and Software Technology}}
  \bibinfo{volume}{51}, \bibinfo{number}{12} (\bibinfo{date}{December}
  \bibinfo{year}{2009}), \bibinfo{pages}{1631--1645}.
\newblock


\bibitem[Nielsen(2012)]%
        {Nielsen_2012}
\bibfield{author}{\bibinfo{person}{Jakob Nielsen}.}
  \bibinfo{year}{2012}\natexlab{}.
\newblock \bibinfo{title}{Usability 101: Introduction to Usability}.
\newblock
  \bibinfo{howpublished}{\url{https://www.nngroup.com/articles/usability-101-introduction-to-usability/}}.
\newblock


\bibitem[Okta(2021)]%
        {Okta_2021}
\bibfield{author}{\bibinfo{person}{Okta}.} \bibinfo{year}{2021}\natexlab{}.
\newblock \bibinfo{title}{HTTP Error 431: Definition, Status, Causes \&
  Solutions}.
\newblock
  \bibinfo{howpublished}{\url{https://www.okta.com/identity-101/http-error-431/}}.
\newblock


\bibitem[Petersen et~al\mbox{.}(2015)]%
        {Petersen_2015}
\bibfield{author}{\bibinfo{person}{Kai Petersen}, \bibinfo{person}{Sairam
  Vakkalanka}, {and} \bibinfo{person}{Ludwik Kuzniarz}.}
  \bibinfo{year}{2015}\natexlab{}.
\newblock \showarticletitle{Guidelines for conducting systematic mapping
  studies in software engineering: An update}.
\newblock \bibinfo{journal}{\emph{Information and Software Technology}}
  \bibinfo{volume}{64} (\bibinfo{date}{August} \bibinfo{year}{2015}),
  \bibinfo{pages}{1--18}.
\newblock


\bibitem[Ri\c{n}\c{k}evi\v{c}s and Torkar(2013)]%
        {Rinkevics_2013}
\bibfield{author}{\bibinfo{person}{K. Ri\c{n}\c{k}evi\v{c}s} {and}
  \bibinfo{person}{R. Torkar}.} \bibinfo{year}{2013}\natexlab{}.
\newblock \showarticletitle{Equality in cumulative voting: A systematic review
  with an improvement proposal}.
\newblock \bibinfo{journal}{\emph{Information and Software Technology}}
  \bibinfo{volume}{55}, \bibinfo{number}{2} (\bibinfo{date}{February}
  \bibinfo{year}{2013}), \bibinfo{pages}{267--287}.
\newblock


\bibitem[Santos et~al\mbox{.}(2018)]%
        {Santos_2018}
\bibfield{author}{\bibinfo{person}{Jos{\'e} Amancio~M. Santos},
  \bibinfo{person}{Jo{\~a}o~B. Rocha-Junior}, \bibinfo{person}{Luciana
  Carla~Lins Prates}, \bibinfo{person}{Rogeres~Santos do Nascimento},
  \bibinfo{person}{Mydi{\~a}~Falc{\~a}o Freitas}, {and}
  \bibinfo{person}{Manoel~Gomes de Mendon\c{c}a}.}
  \bibinfo{year}{2018}\natexlab{}.
\newblock \showarticletitle{A systematic review on the code smell effect}.
\newblock \bibinfo{journal}{\emph{Journal of Systems and Software}}
  \bibinfo{volume}{144} (\bibinfo{date}{October} \bibinfo{year}{2018}),
  \bibinfo{pages}{450--477}.
\newblock


\bibitem[ScienceDirect(2021)]%
        {ScienceDirect_2021_stopwords}
\bibfield{author}{\bibinfo{person}{ScienceDirect}.}
  \bibinfo{year}{2021}\natexlab{}.
\newblock \bibinfo{title}{How do I use the advanced search?}
\newblock
  \bibinfo{howpublished}{\url{https://service.elsevier.com/app/answers/detail/a_id/25974/supporthub/sciencedirect/}}.
\newblock


\bibitem[Shakeel et~al\mbox{.}(2018)]%
        {Shakeel_2018}
\bibfield{author}{\bibinfo{person}{Yusra Shakeel}, \bibinfo{person}{Jacob
  Kr{\"u}ger}, \bibinfo{person}{Ivonne von Nostitz-Wallwitz},
  \bibinfo{person}{Christian Lausberger}, \bibinfo{person}{Gabriel~Campero
  Durand}, \bibinfo{person}{Gunter Saake}, {and} \bibinfo{person}{Thomas
  Leich}.} \bibinfo{year}{2018}\natexlab{}.
\newblock \showarticletitle{(Automated) Literature Analysis - Threats and
  Experiences}. In \bibinfo{booktitle}{\emph{Proceedings of the 13th IEEE/ACM
  International Workshop on Software Engineering for Science (SE4Science
  2018)}}. \bibinfo{publisher}{ACM Press}, \bibinfo{address}{Gothenburg,
  Sweden}, \bibinfo{pages}{20--27}.
\newblock


\bibitem[Shakeel et~al\mbox{.}(2019)]%
        {Shakeel_2019}
\bibfield{author}{\bibinfo{person}{Yusra Shakeel}, \bibinfo{person}{Jacob
  Kr{\"u}ger}, \bibinfo{person}{Ivonne von Nostitz-Wallwitz},
  \bibinfo{person}{Gunter Saake}, {and} \bibinfo{person}{Thomas Leich}.}
  \bibinfo{year}{2019}\natexlab{}.
\newblock \showarticletitle{Automated Selection and Quality Assessment of
  Primary Studies: A Systematic Literature Review}.
\newblock \bibinfo{journal}{\emph{ACM Journal of Data and Information Quality}}
  \bibinfo{volume}{12}, \bibinfo{number}{1} (\bibinfo{date}{November}
  \bibinfo{year}{2019}).
\newblock
\newblock
\shownote{Article 4}.


\bibitem[Singh and Singh(2017)]%
        {Singh_Singh_2017}
\bibfield{author}{\bibinfo{person}{Paramvir Singh} {and}
  \bibinfo{person}{Karanpreet Singh}.} \bibinfo{year}{2017}\natexlab{}.
\newblock \showarticletitle{Exploring Automatic Search in Digital Libraries - A
  Caution Guide for Systematic Reviewers}. In
  \bibinfo{booktitle}{\emph{Proceedings of the 21st International Conference on
  Evaluation and Assessment in Software Engineering (EASE 2017)}}.
  \bibinfo{publisher}{ACM Press}, \bibinfo{address}{Karlskrona, Sweden},
  \bibinfo{pages}{236--241}.
\newblock


\bibitem[Soomro et~al\mbox{.}(2016)]%
        {Soomro_2016}
\bibfield{author}{\bibinfo{person}{Arjumand~Bano Soomro},
  \bibinfo{person}{Norsaremah Salleh}, \bibinfo{person}{Emilia Mendes},
  \bibinfo{person}{John Grundy}, \bibinfo{person}{Giles Burch}, {and}
  \bibinfo{person}{Azlin Nordin}.} \bibinfo{year}{2016}\natexlab{}.
\newblock \showarticletitle{The effect of software engineers' personality
  traits on team climate and performance: A Systematic Literature Review}.
\newblock \bibinfo{journal}{\emph{Information and Software Technology}}
  \bibinfo{volume}{73} (\bibinfo{date}{May} \bibinfo{year}{2016}),
  \bibinfo{pages}{52--65}.
\newblock


\bibitem[Vaughan(2004)]%
        {Vaughan_2004}
\bibfield{author}{\bibinfo{person}{Liwen Vaughan}.}
  \bibinfo{year}{2004}\natexlab{}.
\newblock \showarticletitle{New measurements for search engine evaluation
  proposed and tested}.
\newblock \bibinfo{journal}{\emph{Information Processing and Management}}
  \bibinfo{volume}{40}, \bibinfo{number}{4} (\bibinfo{date}{July}
  \bibinfo{year}{2004}), \bibinfo{pages}{677--691}.
\newblock


\bibitem[Vrana(2007)]%
        {Vrana_2007}
\bibfield{author}{\bibinfo{person}{Radovan Vrana}.}
  \bibinfo{year}{2007}\natexlab{}.
\newblock \showarticletitle{The Importance of Usability in Development of
  Digital Libraries}. In \bibinfo{booktitle}{\emph{Proceedings of the 1st
  International Conference The Future of Information Sciences (INFuture2007)}}.
  \bibinfo{address}{Zagreb, Croatia}, \bibinfo{pages}{379--387}.
\newblock


\bibitem[Xie(2008)]%
        {Xie_2008}
\bibfield{author}{\bibinfo{person}{Hong~Iris Xie}.}
  \bibinfo{year}{2008}\natexlab{}.
\newblock \showarticletitle{Users' evaluation of digital libraries ({DLs}):
  Their uses, their criteria, and their assessment}.
\newblock \bibinfo{journal}{\emph{Information Processing and Management}}
  \bibinfo{volume}{44}, \bibinfo{number}{3} (\bibinfo{date}{May}
  \bibinfo{year}{2008}), \bibinfo{pages}{1346--1373}.
\newblock


\bibitem[Yang et~al\mbox{.}(2016)]%
        {Yang_Liang_2016}
\bibfield{author}{\bibinfo{person}{Chen Yang}, \bibinfo{person}{Peng Liang},
  {and} \bibinfo{person}{Paris Avgeriou}.} \bibinfo{year}{2016}\natexlab{}.
\newblock \showarticletitle{A systematic mapping study on the combination of
  software architecture and agile development}.
\newblock \bibinfo{journal}{\emph{Journal of Systems and Software}}
  \bibinfo{volume}{111} (\bibinfo{date}{January} \bibinfo{year}{2016}),
  \bibinfo{pages}{157--184}.
\newblock


\end{thebibliography}


\end{document}